\documentclass[]{emulateapj}

\usepackage{natbib}
\usepackage{bm}
\usepackage{amsmath}
\usepackage{txfonts}
\usepackage{epstopdf}
\usepackage{graphicx,color}

\newcommand{\rp}{R_{\rm p}}
\newcommand{\rH}{r_{\rm H}}

\newcommand{\rs}{r_{\rm s}}

\newcommand{\Mp}{M_{\rm p}}

\newcommand{\secref}[1]{Section \ref{#1}}
\newcommand{\figref}[1]{Figure \ref{#1}}

\shorttitle{3D Isentropic Torque}
\shortauthors{Fung, Masset, Lega \& Velasco}

\begin{document}

\title{Planetary Torque in 3D Isentropic Disks}
\author{Jeffrey Fung\altaffilmark{1,2}, Fr\'{e}d\'{e}ric Masset\altaffilmark{3}, Elena Lega\altaffilmark{4}, and David Velasco\altaffilmark{3}}

\altaffiltext{1}{Department of Astronomy, University of California at Berkeley, Campbell Hall, Berkeley, CA 94720-3411}
\altaffiltext{2}{NASA Sagan Fellow}
\altaffiltext{3}{Instituto de Ciencias F\'\i sicas, Universidad Nacional Aut\'onoma de M\'exico, Av. Universidad s/n, 62210 Cuernavaca, Mor., Mexico}
\altaffiltext{4}{Universit\'e de la C\^ote d'Azur, Observatoire de la  C\^ote d'Azur, CNRS, Laboratoire Lagrange UMR 7293. }

\email{email: jeffrey.fung@berkeley.edu}

\begin{abstract}
Planet migration is inherently a three-dimensional (3D) problem, because Earth-size planetary cores are deeply embedded in protoplanetary disks. Simulations of these 3D disks remain challenging due to the steep requirement in resolution. Using two different hydrodynamics code, \texttt{FARGO3D} and \texttt{PEnGUIn}, we simulate disk-planet interaction for a 1 to 5 Earth-mass planet embedded in an isentropic disk. We measure the torque on the planet and ensure that the measurements are converged both in resolution and between the two codes. We find that the torque is independent of the smoothing length of the planet's potential ($\rs$), and that it has a weak dependence on the adiabatic index of the gaseous disk ($\gamma$). The torque values correspond to an inward migration rate qualitatively similar to previous linear calculations. We perform additional simulations with explicit radiative transfer using \texttt{FARGOCA}, and again find agreement between 3D simulations and existing torque formulae. We also present the flow pattern around the planets, and show that active flow is present within the planet's Hill sphere, and meridional vortices are shed downstream. The vertical flow speed near the planet is faster for a smaller $\rs$ or $\gamma$, up to supersonic speeds for the smallest $\rs$ and $\gamma$ in our study. 
\end{abstract}

\keywords{accretion, accretion disks --- methods: numerical --- planets and satellites: formation --- protoplanetary disks --- planet-disk interactions --- circumstellar matter --- stars: variables: T Tauri, Herbig Ae/Be}

\section{Introduction}
\label{sec:intro}
Newly born planets interact gravitationally with their natal circumstellar disks. As the planet's tidal force
exerts a torque on the disk, the back-reaction from the disk also torques
the planet, causing it to migrate. 
The study of planet migration is one of the links that connects the initial formation 
of planets to their final positions in their planetary systems, and is therefore
essential to explaining the statistical distribution of planets \citep[e.g.,][]{Petigura13,Silburt15,Dressing15}.
Calculations of planet migration rates
have been done extensively for planets interacting with razor-thin, two-dimensional (2D)
disks (for a review, see \citet{Baruteau13} and references therein), and, 
to a lesser extent, with more realistic, three-dimensional disks (3D) 
\citep[e.g.][]{Tanaka02,Bate03,DAngelo03,Masset06,DAngelo10,Uribe11,Lega14,Benitez15}.

Recently, \citet{Fung15} (hereafter FAW15) reported a torque on an Earth-size
planet embedded in an isothermal disk that significantly deviates from 
linear estimates, such as those by \citet{Tanaka02}. This opens the possibility 
that the co-orbital region may behave in a way previously unanticipated, thus 
generating a new component in the torque.
Their results, however, also show insufficient resolution near the planet, 
leading to the need to exclude the planet's vicinity for their torque 
calculation. The existence of this new torque component therefore 
demands verification.

Also demonstrated by recent results in the literature, including those of 
\citet{Benitez15} and FAW15, is that in 3D disks, the density structure close 
to the planet, within the scale of the planet's Hill radius, can potentially
have an overwhelming influence on the planetary torque. It is therefore essential 
that the thermal properties of the gas in this area be properly modelled.
Protoplanetary disks models commonly assume a locally isothermal structure.
While this may be appropriate for the global disk, it is unsuitable for the gas 
surrounding the planet. In reality, the planet's atmosphere is 
opaque to its own thermal emission and cools less efficiently than the disk, and so
it should more resemble a heated envelope, within 
which convection is expected to operate, generating a constant entropy profile. 
This implies an isentropic equation of state may be more applicable to the density
structure close to the planet. Additionally, the hydrostatic profile of an 
isothermal gas inside the planet's gravitational potential is exponential, which 
makes achieving numerical convergence challenging. As we will show later in this paper, 
convergence is more readily achieved when the adiabatic index, $\gamma$, is larger.

This paper aims to obtain a converged measurement of the torque on an
Earth-size planet embedded in a 3D isentropic disk. We will investigate possible
dependencies on both the adiabatic index $\gamma$, and the smoothing 
length of the planet's gravitational potential $\rs$.
To identify potential code bias, we simulate identical models with two 
hydrodynamics codes: \texttt{FARGO3D} and \texttt{PEnGUIn}. To further identify 
possible discrepancies between 2D and 3D disks, we also present
additional simulations with explicit radiative transfer using \texttt{FARGOCA}.

\secref{sec:numerics} contains code descriptions and simulation
parameters. \secref{sec:results} presents our results for both the torque and the flow 
structure around the planet. \secref{sec:conclude} concludes and discusses implications
of our results.

\section{Numerical Method}
\label{sec:numerics}
The simulations are performed in spherical coordinates, where $r$, $\phi$, and $\theta$ denote 
the usual radial, azimuthal, and polar coordinates. For convenience, we also denote $R\equiv r\sin{\theta}$
and $z\equiv r\cos{\theta}$ as the cylindrical radial and vertical coordinates.
\texttt{FARGO3D} and \texttt{PEnGUIn} solve the compressible Euler equations:
\begin{equation}
\label{eqn:cont_eqn}
\frac{{\rm D}\rho}{{\rm D} t} = -\rho\left(\bm{\nabla} \cdot\bm{v}\right) \,,
\end{equation}
\begin{equation}
\label{eqn:moment_eqn}
\frac{{\rm D} \bm{v}}{{\rm D} t} = -\frac{1}{\rho}\bm{\nabla} p - \bm{\nabla} \Phi \,,
\end{equation}
where ${\rm D}/{\rm D}t$ is the Lagrangian derivative, $\rho$ is the density, 
$p$ the gas pressure, $\bm{v}$ the velocity, and $\Phi$ the gravitational potential 
of the central star and the planet, representing the only external force we consider
in this study.

For our model with explicit radiative transfer, we include the following energy equation:
\begin{equation}
\label{eqn:energy_eqn}
\frac{{\rm D} e}{{\rm D} t} = -e\left(\bm{\nabla} \cdot\bm{v}\right)
-p \left(\bm{\nabla} \cdot \bm {v}\right) - \bm{\nabla} \cdot \bm F + Q^+\,,
\end{equation}
where $e$ is the thermal energy density: $e=\rho c_v T$, $c_v$ the
  specific heat at constant volume and $T$ the gas temperature. The
  term $p(\bm{\nabla} \cdot \bm v)$ represents compressional heating, 
$Q^+ = (\mathbb{T} \nabla) \cdot \bm v$ viscous heating (with
  $\mathbb{T}$  the viscous stress tensor, see for example
  \citep{Mihalas}) and  $\bm F$ is the radiative flux. We have chosen
  a disk with a constant viscosity \footnote{For the radiative disk the
    r.h.s of
    Eq.\ref{eqn:moment_eqn} has an additional term corresponding to
    the divergence of the viscous stress tensor ($-\frac{1}{\rho}\bm
    \nabla \cdot \mathbb{T}$)}$\nu=10^{-5} \rp^2 \Omega_{\rm p}$, where $\rp$ is the orbital radius of the planet's orbit, and $\Omega_{\rm p}$ the planet's orbital frequency.
The radiative flux is calculated in the framework of the
  flux-limited diffusion approximation (\citet{Levermore81}), 
\begin{align}
  \bm F  = -\frac{c f_\lambda 4 a_R T^3}{\rho \kappa}  \nabla T
\end{align}
where $f_\lambda$ is a flux-limiter \citep{Kley89}, $c$ is the
speed of  light, $a_R$ the radiation constant and $\kappa$ the
Rosseland mean opacity, set to be 0.01 $\rm cm^2 g^{-1}$ (or $3.28\times 10^3$in code units).

We note that for our isentropic models where viscous heating is not a concern (unlike the radiative case), we have delibrately chosen not to include viscosity, because we consider it a more physical choice. Rather than being viscous, real protoplanetary disks are turbulent, and it is unclear that this turbulence translates to viscosity when considering detail flow structures on the scale of the planet's Hill radius. In fact, viscosity could easily generate unrealistic behavior, such as preventing cirumplanetary flow by removing angular momentum too rapidly.

The simulations are performed in a frame centered on the star,
and we fix the planet on a circular orbit in the disk midplane. 
$\Phi$ is therefore:
\begin{equation}\label{eqn:potential}
\Phi = -\frac{GM}{1+q}\left[\frac{1}{r} + \frac{q}{\sqrt{r^2 + \rp^2 - 2R\rp\cos{\phi'} + \rs^2}} - \frac{qR\cos{\phi'}}{\rp^2}\right] \,,
\end{equation}
where $G$ is the gravitational constant, $M=M_* + \Mp$ the total mass of the star and the planet, $q = \Mp / M_*$ the planet-to-star mass ratio, $\rs$ the smoothing length of the planet's potential, and $\phi' = \phi-\phi_{\rm p}$ denotes the azimuthal separation from the planet. We set $GM=1$ and $\rp=1$, so that the Keplerian velocity and frequency $v_{\rm k} = \sqrt{GM/r}$ and $\Omega_{\rm k} = \sqrt{GM/r^3}$ both equal 1 at the planet's orbit. We also label the planet's orbital speed as $v_{\rm p}$ for convenience. The third term in the bracket is the indirect potential due to the non-inertial frame. 

For comparison with FAW15, we choose $q=1.5\times10^{-5}$, or about 5 Earth masses for a Solar-mass star, for most of our simulations, as well as an additional few with $q=3\times10^{-6}$, or about 1 Earth mass. The Hill radius of a planet is:
\begin{equation}
\label{eqn:rH}
\rH = \rp\left(\frac{q}{3}\right)^{\frac{1}{3}} \,,
\end{equation}
which comes to $\rH\approx0.017\rp$ for a 5 Earth-mass ($M_\oplus$) planet. We set $\rs$ to be a small fraction of $\rH$, and look into torque's dependence on it. Our choices for $\rs$ are between 3\% to 10\% of $\rH$. Realistically, $\rs$ should correspond to the physical size of the planet, which, for example, is about 0.4\% of $\rH$ for the Earth. It is however not numerically feasible to use such a small $\rs$, as the resolution required would be prohibitively high.

Completing our set of equations is the isentropic equation of state:
\begin{equation}
\label{eqn:isen_eqn}
p = \frac{c_0^2 \rho_0}{\gamma}  \left(\frac{\rho}{\rho_0}\right)^{\gamma} \,,
\end{equation}
where $c_0$ is the adiabatic sound speed when $\rho = \rho_0$, and the
normalization $\rho_0$ is set to 1. The sound speed is calculated 
from $c_s = \sqrt{\gamma p/\rho}$. In all of our simulations, we fix
$c_0=0.03$, and scale our density profile such that $\rho = \rho_0$ at
the planet's location. As a result, the sound speed near the planet is
always 3\% of the local Keplerian speed, regardless of $\gamma$. This
is the same sound speed as the one used by FAW15. We simulate disks
with 3 different values of $\gamma$: $1.2$, $1.4$, and $1.67$.

For the model with radiative transfer we instead use the ideal gas equation of state:
\begin{equation}
\label{eqn:rad_eqn}
p = \frac{R_{gas} \rho T}{\mu} 
\end{equation}
with mean molecular weight $\mu$, $\mu=2.3$, and gas constant $R_{gas}$. 
The sound speed is calculated using $\gamma=1.4$. The dependence of 
the torque on $\gamma$ for radiative models was previously
investigated by \citet{Bitsch13}.
\subsection{Initial and boundary conditions}
\label{sec:initial}
The disks are initialized assuming hydrostatic equilibrium. The density profile is:
\begin{equation}
\label{eqn:rho_init}
\rho = \rho_0 \left[ \left(\frac{R}{\rp}\right)^{-(\beta+\frac{3}{2})\frac{2(\gamma-1)}{\gamma+1}} - \frac{GM(\gamma-1)}{c_0^2}\left(\frac{1}{R}-\frac{1}{r}\right)\right]^{\frac{1}{\gamma-1}} \,,
\end{equation}
where $\beta$ defines the surface density profile $\Sigma\propto
R^{-\beta}$. In this work we choose $\beta=3/2$. This produces a
constant initial vortensity profile, $(\bm{\nabla}\times\bm{v_{\rm
    k}}) / \Sigma$, which, in 2D disks, would imply a net zero
horseshoe drag \citep{Ward91}. \citet{Masset16} have shown
that the same applies in 3D, as long as vorticity is kept in
the vertical direction (i.e. no vortex tilting). With both a 
constant entropy and vortensity, none of the known sources of 
corotation torque is present in our model.

The orbital frequency of the disk in hydrostatic equilibrium is modified by the radial pressure gradient:
\begin{equation}
\label{eqn:omega}
\Omega = \sqrt{\Omega_{\rm k}^2 + \frac{1}{r\rho}\frac{\partial p}{\partial r}} \,,
\end{equation}
while the radial and polar velocities are zero. The planet is introduced to the disk gradually, where its mass increases to the desired value over the first orbit. 

Our simulation domain spans $0.7\rp$ to $1.3\rp$ in the radial, and the full $2\pi$ in azimuth. Since our density profile is polytropic, it falls to negative values when 
\begin{equation}
\label{eqn:top_limit}
z > z_{\rm lim}\equiv\sqrt{ \frac{2}{\gamma-1} } \frac{c_0}{\Omega_{\rm k}} R^{-(\beta+\frac{3}{2})\frac{\gamma-1}{\gamma+1}} \,.
\end{equation}
We therefore determine the top polar boundary by setting it below $\arctan{(z_{\rm lim}/R)}$ for all $R$ within our simulation domain, which is $0.087$ radian above the midplane when $\gamma=1.2$, $0.066$ when $\gamma=1.4$, and $0.048$ when $\gamma=1.67$. For the radiative simulations, the top polar boundary is placed at $0.1$ radian above the midplane. In all models, we set our bottom boundary at the disk midplane.

For our radial boundaries, we apply a fixed boundary condition where
the disk variables remain at their initial values. An additional wave killing zone \citep{DeValBorro06}
is implemented in the radiative simulations to prevent the reflection of density
waves. The polar boundaries are reflective, both at the top, to prevent mass from entering or leaving the domain, and at the bottom, to ensure the disk is symmetric across the midplane.

\subsection{Code descriptions and simulation resolution}
\label{sec:resolve}

We use three different resolutions to test the numerical convergence of our results; a ``low'' resolution where the planet's Hill radius $\rH$ is resolved by $\sim 30$ cells, a ``medium'' resolution where it is doubled to $\sim 60$ cells, and a ``high'' resolution with again doubled resolution to $\sim 120$ cells. Our codes use different methods to achieve these resolutions in the vicinity of the planet: nested grid in \texttt{FARGO3D}, and non-uniform grid geometry in \texttt{PEnGUIn} and \texttt{FARGOCA}.

In our three codes, the reference frame co-rotates with the planet, but the Coriolis force is not computed as an explicit source term, as shown by Eq.~(\ref{eqn:moment_eqn}). Rather, it is absorbed into the conservative form of the angular momentum equation \citep{Kley98}.

In the following, we give a brief description of these codes.

\subsubsection{\texttt{FARGO3D}}
\label{sec:FARGO3D}
The code \texttt{FARGO3D} \citep{FARGO3D} is here used with a newly implemented nested grid capability. On top of a base grid that has the boundaries specified in section~\ref{sec:initial}, we use a hierarchy of nested grids with a doubling up of the resolution between successive grid levels. The limits of our nested grids are given in Tab.~\ref{tab:nested}. Within each grid, the hydrodynamical solver described by \citet{FARGO3D} is used, with orbital advection deactivated. Boundary conditions of the nested meshes are imposed by performing a trilinear interpolation of the underlying coarser mesh in a three-cell wide layer of ghost zones surrounding the mesh, except for the boundaries at the midplane ($\theta=\pi/2$) which are reflective. Upon integration on a given level (hereafter fine level), the information is communicated to the underlying coarser level (hereafter coarse level) in two ways: (i) in the outermost two-cell wide contour of the coarse level covered by the fine one, the different hydrodynamics quantities are replaced by their averaged values from the fine level and (ii) the fluxes of mass and momentum on the contour of the fine level are used to update the quantities on the coarse level, which ensures that the code conserves mass and angular momentum to machine accuracy. The integration is done recursively across the whole hierarchy of nested grids. If a given level is advanced in time with a time step $\Delta t$, the next finer level is either also advanced over a time step $\Delta t$, or it is advanced twice with a time step $\Delta t/2$. Which of these two possibilities is chosen depends on which one yields the largest advance in time for a given computational cost. Our implementation runs on Graphics Processing Units (GPUs) and is parallelized using MPI (\emph{Message Passing Interface}). The details of our implementations will be presented elsewhere.

\begin{deluxetable}{lllllll}
\tablecaption{\label{tab:nested} Position of the nested grids}
\tablehead{$\phi_\mathrm{min}$&$r_\mathrm{min}/R_p$&$\theta_\mathrm{min}$&$\phi_\mathrm{max}$&$r_\mathrm{max}/R_p$&$\theta_\mathrm{max}$&level}
\startdata 
$-1.0$&$0.89$&$1.54$&$1.0$&$1.11$&$\pi/2$&$1$\\
$-0.2$&$0.91$&$1.545$&$0.2$&$1.09$&$\pi/2$&$2$\\
$-0.1$&$0.95$&$1.553$&$0.1$&$1.05$&$\pi/2$&$3$\\
$-0.04$&$0.97$&$1.556$&$0.04$&$1.03$&$\pi/2$&$4$\\
$-0.025$&$0.98$&$1.559$&$0.025$&$1.02$&$\pi/2$&$5$\\
$-0.018$&$0.985$&$1.562$&$0.018$&$1.015$&$\pi/2$&$6$\\
\enddata
\tablecomments{Levels 1 to 4 are used in the ``low''
  resolution runs, 1 to 5 in the ``medium'' resolution runs, and 1 to
  6 in the ``high'' resolution runs. This table lists the intended
  limits, set by the user. The actual limits differ slightly, as they
  have to fall on the interfaces between cells of the coarser level.}
\end{deluxetable}

\subsubsection{\texttt{PEnGUIn}}
\label{sec:PEnGUIn}
\texttt{PEnGUIn} ({\bf P}iecewise Parabolic Hydro-code {\bf En}hanced
with {\bf G}raphics Processing {\bf U}nit {\bf I}mplementatio{\bf n})
is a Lagrangian, dimensionally-split, shock-capturing
hydrodynamics code that runs on GPUs
\citep{MyThesis}.

Different from \texttt{FARGO3D}, \texttt{PEnGUIn} achieves high resolution using a 
non-uniform grid geometry. We assign small cell sizes near the planet, and 
continuously increase the cell size as the distance from the planet grows. The
prescription is as follows:
\begin{align}
\label{eqn:penguin_res}
\Delta = &\frac{x_{\rm max}^a}{N a} x^{1-a} + \Delta_{\rm min} \,, \\
a = &\frac{x_{\rm max}}{N \Delta_{\rm max}} \,,
\end{align}
where $\Delta$ is the cell size, $x$ is the distance away from the planet, $x_{\rm max}$ is the maximum value of $x$ (i.e. the domain boundary), $N$ is the number of cells between $0$ and $x_{\rm max}$, and $\Delta_{\rm min}$ and $\Delta_{\rm max}$ are the assigned cell sizes at $x=0$ and $x=x_{\rm max}$ respectively. For our low, medium, and high resolution runs, we set $\Delta_{\rm min}=5.0\times10^{-4}$, $2.5\times10^{-4}$, and $1.25\times10^{-4}$ respectively. $\Delta_{\rm max}$ is always fixed at $0.002$ in the radial and polar directions, and $0.015$ in the azimuthal direction.

\subsubsection{\texttt{FARGOCA}}
\label{sec:FARGOCA}
The code FARGOCA (FARGO with {\bf C}olatitude {\bf A}dded)
 \citep{Lega14}  is
based on  the FARGO code \citep{Masset00} extended  to 3 dimensions
with the additional  introduction of an energy equation to provide a
realistic modeling of radiative effects \citep{Kley09}.
The fluid equations are solved using finite-differences with a
time-explicit-implicit multistep procedure. Precisely, concerning the
energy equation, we first update the energy by explicit integration of
the compressional heating term and in a separate step we integrate 
the  viscous heating and  the radiative diffusion terms. In this second step
we follow the backward Euler method, which is an
unconditionally stable implicit method solved with a standard SOR
(Successive Overrelaxation Reduction) solver. 
The code is parallelized using a hybrid combination of MPI between the
nodes and of OpenMP on shared memory multi-core processors. 
High resolution is achieved as in \texttt{PEnGUIn} using the nonuniform
grid geometry with the prescription detailed above.
Since hydrodynamical 3D calculations
of radiative disks are very expensive in computational time 
we have  moderately lower resolution   with respect to the
values used for isentropic disks with both 
\texttt{FARGO3D} and \texttt{PEnGUIn} .
We  set in the following for our low, medium, and high resolution
radiative runs: $\Delta_{\rm min}=1.2\times10^{-3}$, $6\times10^{-4}$, and $3\times10^{-4}$ respectively.


\section{Results}
\label{sec:results}
\subsection{Planetary Torque}
\label{sec:torque}

\begin{figure}[]
\includegraphics[width=0.99\columnwidth]{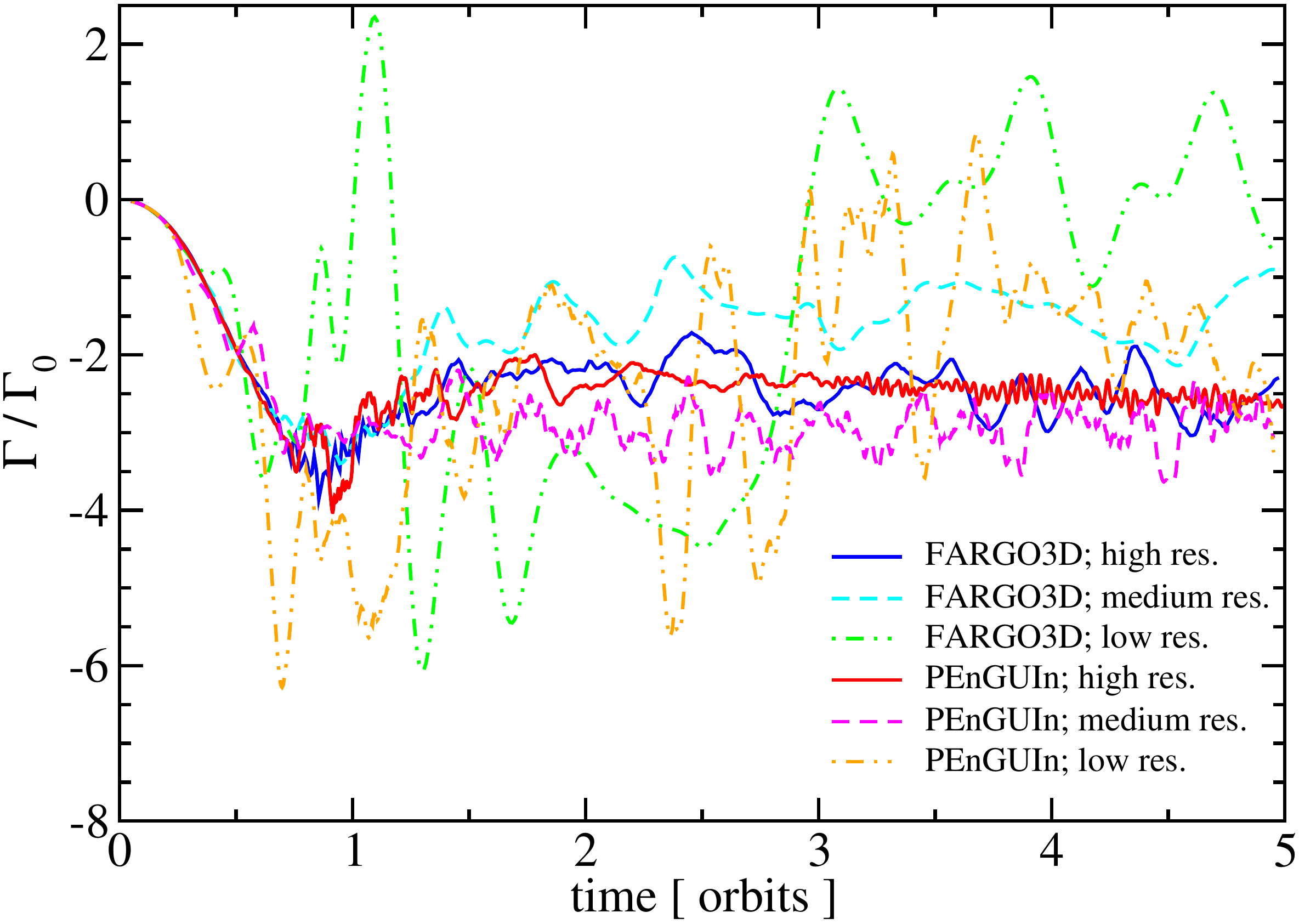}
\caption{Torque on a 5 $M_\oplus$ planet as a function of time for different resolutions with both \texttt{FARGO3D} and \texttt{PEnGUIn}. All models shown have $\{\rs/\rH,\gamma\}=\{0.03, 1.2\}$. Solid, dashed, and dash-dot-dot lines corresponds to ``high'', ``medium'', and ``low'' resolution defined in \secref{sec:resolve}. \texttt{FARGO3D} simulations are in blue, cyan and green, while \texttt{PEnGUIn} ones are in red, magenta, and orange. Data points are time-averaged over 0.1 orbit. At high resolution, both codes converge to the same torque value after 5 orbits, $\sim-2.5\Gamma_0$, and the agreement between them is within $5\%$.}
\label{fig:conv}
\end{figure}

\begin{figure*}[]
\includegraphics[width=1.99\columnwidth]{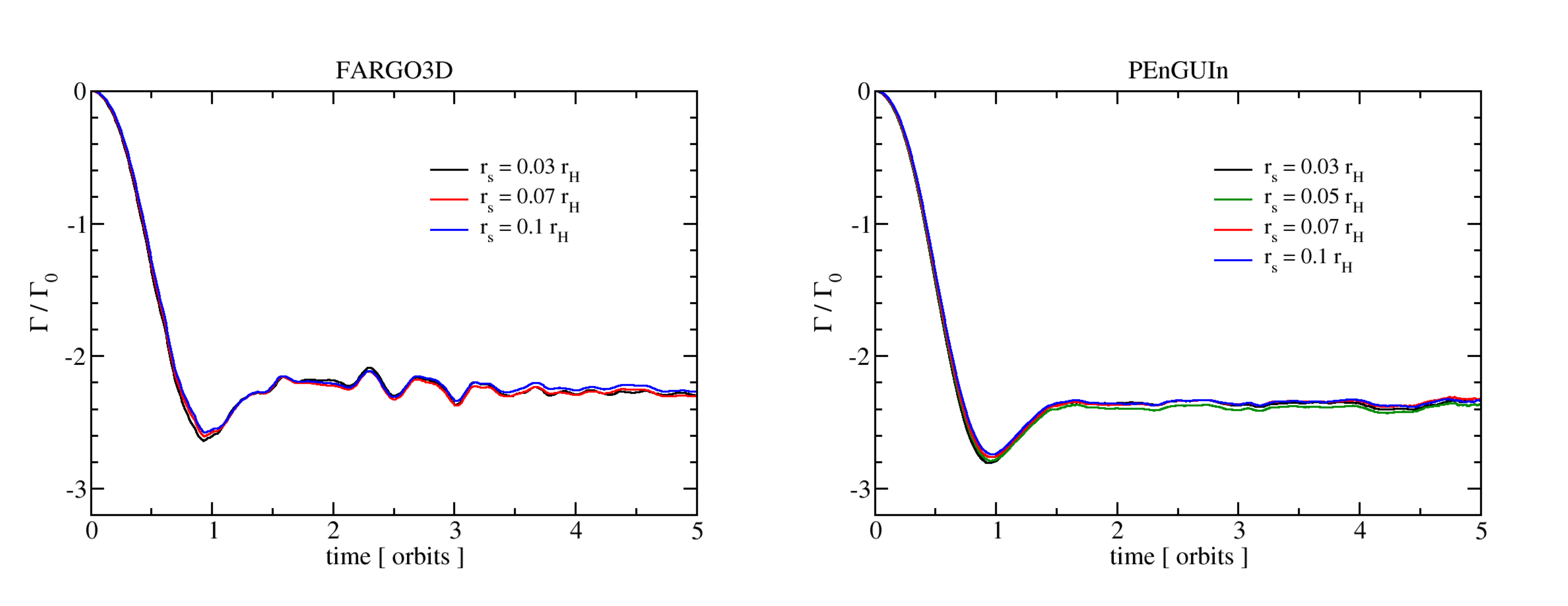}
\caption{Torque on a 5 $M_\oplus$ planet as a function of time for different values of the smoothing length $\rs$. The left panel plots simulation results from \texttt{FARGO3D}, and right panel from \texttt{PEnGUIn}. All models shown have $\gamma$=1.4. Overall we find the torque to be independent of $\rs$.}
\label{fig:smooth}
\end{figure*}

\begin{figure*}[]
\includegraphics[width=1.99\columnwidth]{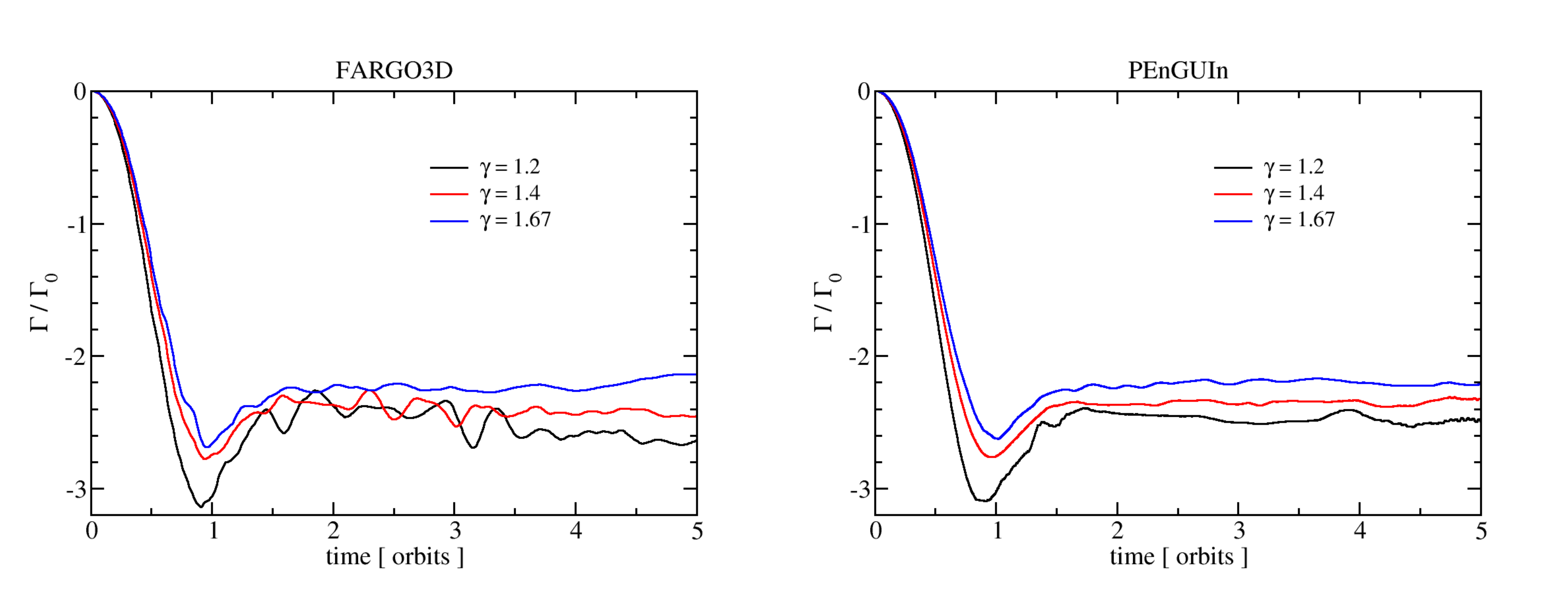}
\caption{Torque on a 5 $M_\oplus$ planet as a function of time for different values of the adiabatic index $\gamma$. All models shown have $\rs=0.07\rH$. Both codes show that the torque has a weak dependence on $\gamma$.}
\label{fig:gam}
\end{figure*}

\begin{figure}[]
\includegraphics[width=0.99\columnwidth]{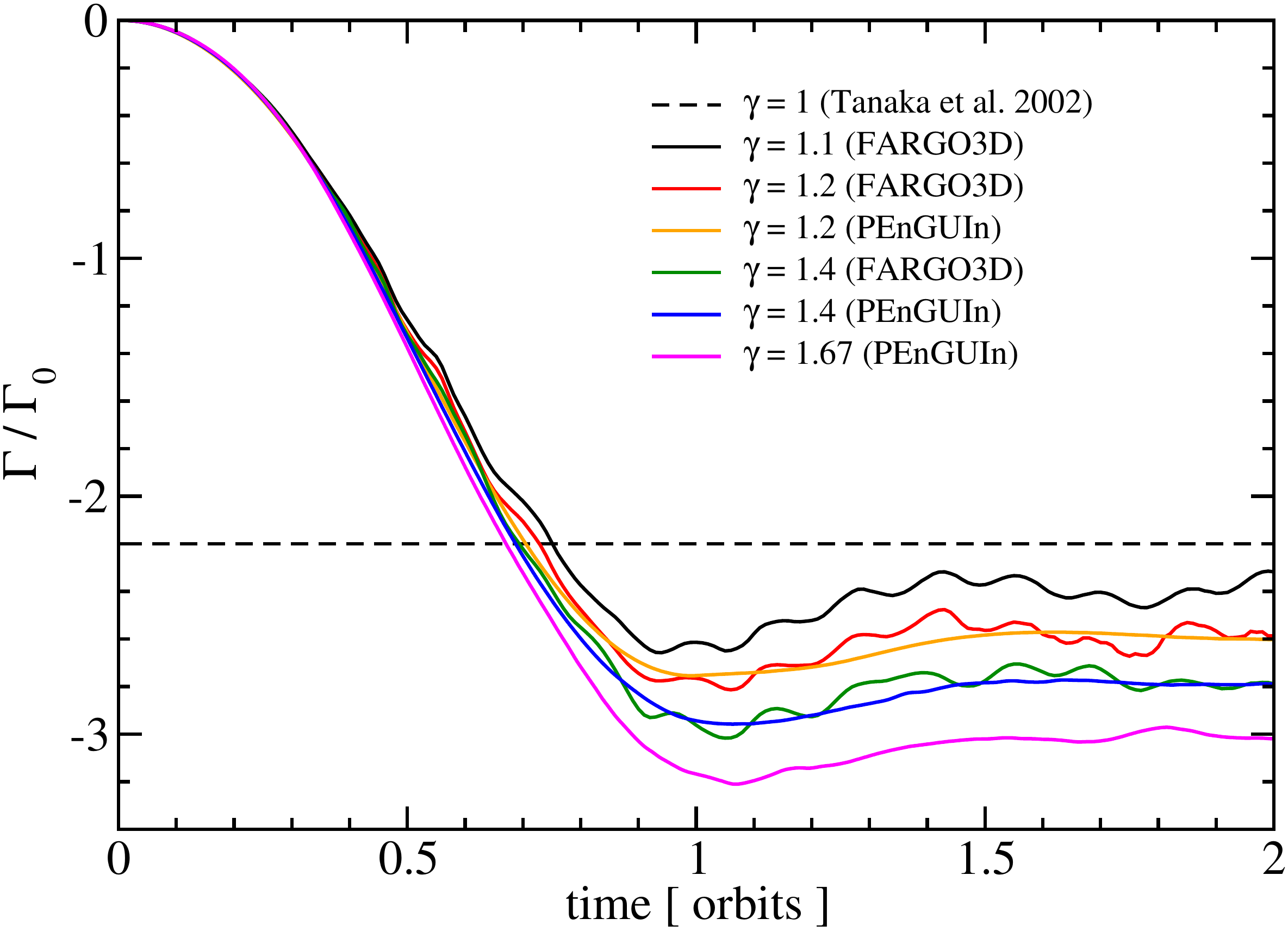}
\caption{Similar to \figref{fig:gam}, but for 1 $M_\oplus$ planets. All models shown have $\rs=0.07\rH$. The dashed line indicates the linear torque computed by \citet{Tanaka02} for an isothermal disk. Comparing to \figref{fig:gam}, we find the trend in $\gamma$ is reversed, and is compatible with linear estimates in the isothermal limit.}
\label{fig:1ME}
\end{figure}

\begin{figure*}[]
\includegraphics[width=1.99\columnwidth]{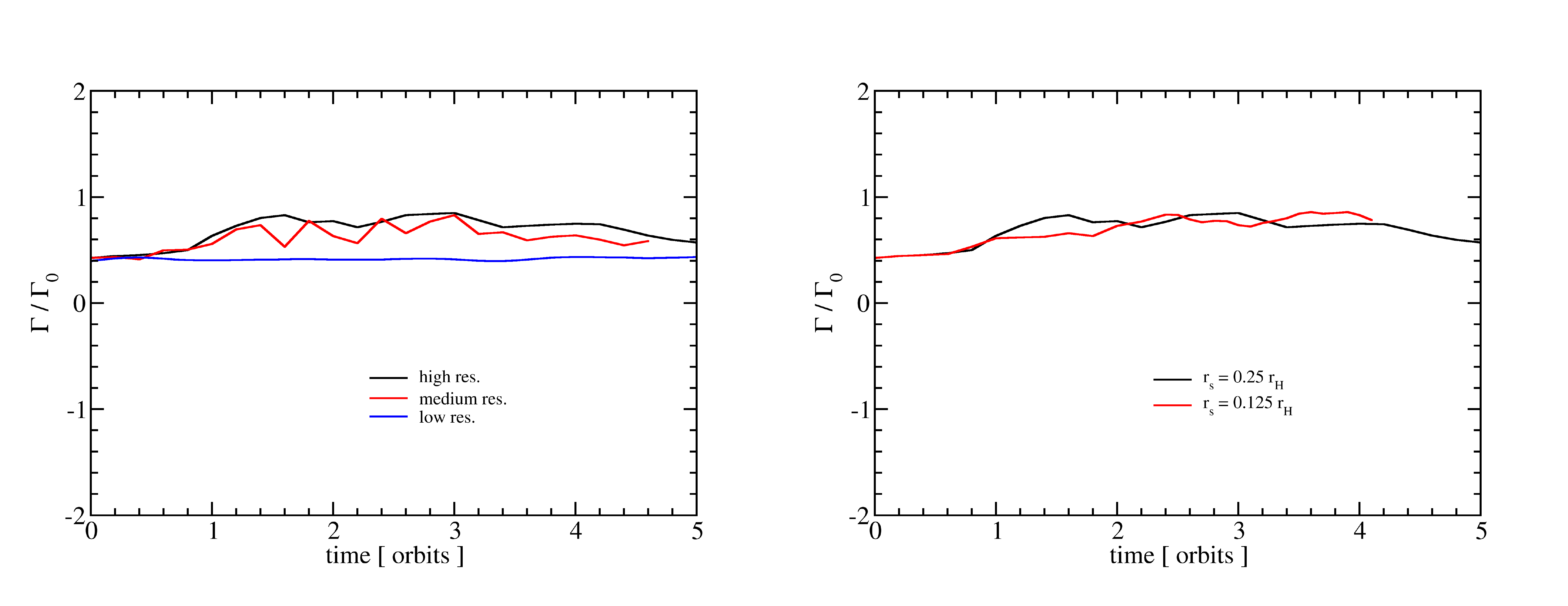}
\caption{Torque on a 5 $M_\oplus$ planet as a function of time, for the radiative models from \texttt{FARGOCA}. Left panel: torque measured for different resolutions values. The ``low'', ``medium'', and ``high'' resolutions for the radiative case are defined in \secref{sec:resolve}. The medium and high resolution runs are restarted from the low resolution run after the torque has come to a steady state (about 20 orbits). We measure a final torque of $\sim 0.55~\Gamma_0$, in agreement with the expected
analytic value. Right panel: torque measured for two different smoothing lengths $\rs$. Similar to \figref{fig:smooth}, we observe no dependence on $\rs$.}
\label{fig:rad}
\end{figure*}

\begin{figure*}[]
\includegraphics[width=1.99\columnwidth]{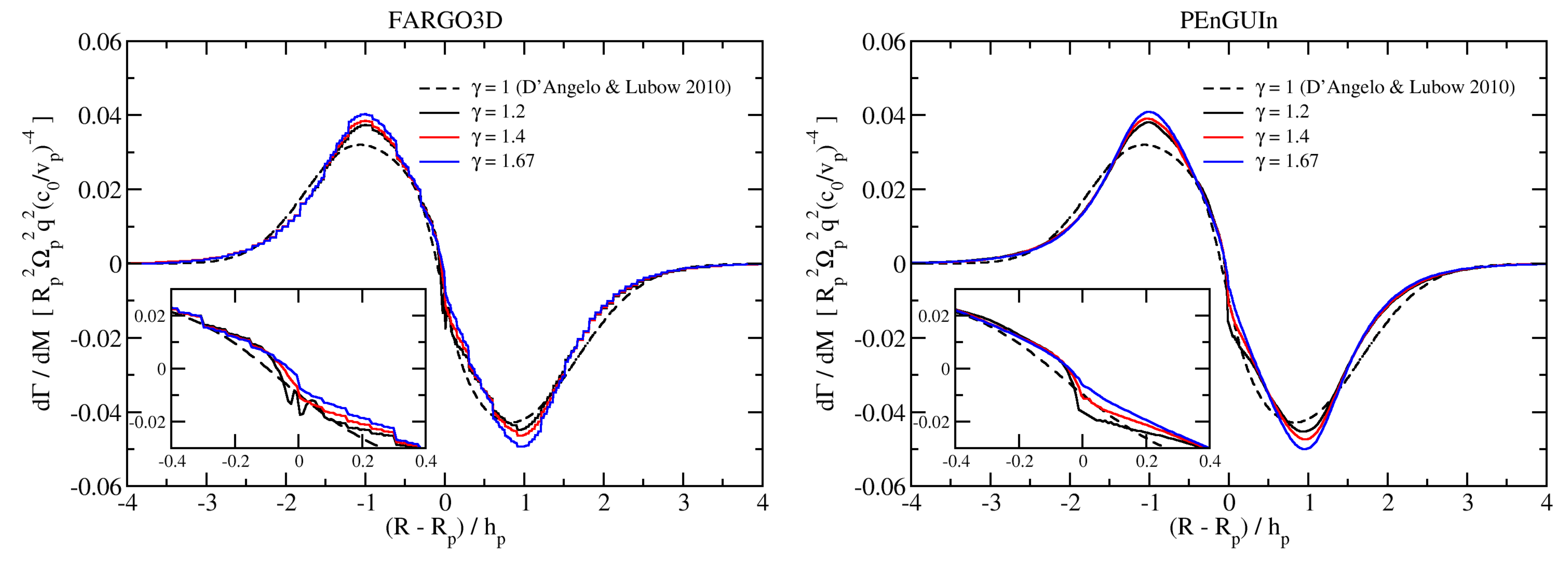}
\caption{Torque density from the same models as those in \figref{fig:gam}, obtained at the end of our simulations. Also shown as the dashed black curves are the isothermal model of \citet{DAngelo10}. The insets zoom in on the vicinity around the planet, revealing a small corotation torque that increases with decreasing $\gamma$. The two codes agree well, with the \text{FARGO3D} result showing more noise coming from the lower resolution regions in the nested mesh.}
\label{fig:tor_den_5ME}
\end{figure*}

\begin{figure}[]
\includegraphics[width=0.99\columnwidth]{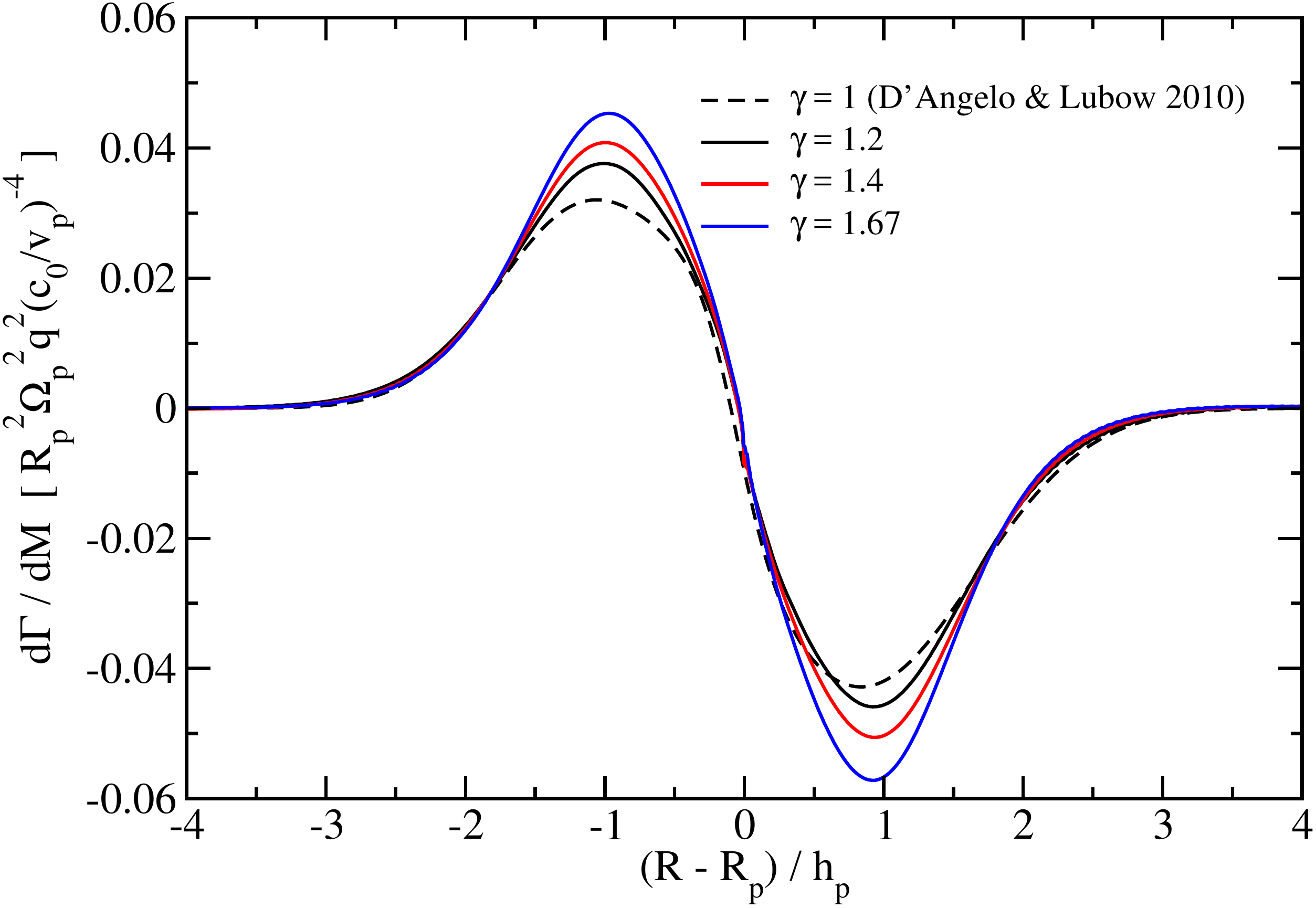}
\caption{Torque density from our 1 $M_\oplus$, $\rs=0.07\rH$ PEnGUIn simulations. Similar to \figref{fig:tor_den_5ME}, we also overlay the isothermal model of \citet{DAngelo10} here as the dashed black curve. We find that the 1 $M_\oplus$ models do not show any corotation torque, and the Lindblad torque is stronger for a larger $\gamma$.}
\label{fig:tor_den_1ME}
\end{figure}

We run each simulation to 5 orbits, at which point a steady torque on
the planet has been established, but the planet has yet to create
nonlinear modifications to the disk structure, such as planetary
gaps. We defer a discussion on the long-term effects to
\secref{sec:conclude}. We measure the net torque on the planet without
excluding any region, and label the torque $\Gamma$, normalizing it by $\Gamma_0 \equiv \Sigma_0 \rp^4 \Omega_{\rm p}^2
q^2 (c_0/v_{\rm p})^{-2}$ 
(this also applies to the radiative model, where the equilibrium temperature at the planet's position has the same sound speed as $c_0$).

Convergence requires higher resolution for smaller $\rs$ or $\gamma$ values. For instance, we find that our low resolution setup is typically sufficient when $\rs=0.1\rH$, but high resolution is needed when $\rs<0.05\rH$. Typically, $\rs$ needs to be resolved by at least 3 cells when $\gamma\gtrsim1.4$, and about 4 to 5 cells when $\gamma=1.2$. We achieved convergence to within a few percent for nearly all of our models, both in resolution for \texttt{FARGO3D} and \texttt{PEnGUIn} individually, and in code comparison between them. \figref{fig:conv} plots the torque of our $\{\rs/\rH,~\gamma\}=\{0.03,~1.2\}$ model, as a function of time. This model has the lowest $\rs$ and $\gamma$ values in our parameter space and shows the highest level of fluctuation; nonetheless, the agreement between \texttt{PEnGUIn} and \texttt{FARGO3D} is within $5\%$. In the rest of this paper, we will only present the converged torque measurements, which are either from medium or high resolution simulations, with the medium resolution results verified by shorter high resolution runs.

In \figref{fig:smooth}, we find that the torque has no dependence on $\rs$ for the range of values we considered ($\rs = 0.03$ to 0.1$~\rH$). The models shown are of $\gamma=1.4$, but the same is found for other $\gamma$ values as well. In \figref{fig:gam}, we plot models with different $\gamma$'s, and observe a weak trend where the torque is more negative for a smaller $\gamma$. In \texttt{PEnGUIn} simulations, we measure $-2.5$, $-2.3$, and $-2.2~\Gamma_0$ for $\gamma =$ 1.2, 1.4, and 1.67 respectively; \texttt{FARGO3D} simulations show the same trend with torques of $-2.6$, $-2.4$, and $-2.2$, respectively. \figref{fig:tor_den_5ME} plots the torque density ${\rm d}\Gamma/{\rm d}M$ (torque per mass at each annulus) of these models, with the 3D isothermal model of \citet{DAngelo10} overlayed (interpolated between their $\Sigma\propto r^{-1}$ and $r^{-2}$ models.). We find a torque arising in the planet's co-orbital region as $\gamma$ decreases, which largely accounts for the observed trend in $\gamma$.

This corotation torque is not predicted by the linear calculations of \citet{Tanaka02}. This is likely because the 5 $M_\oplus$ planet we choose is about half the disk's thermal mass:
\begin{equation}
M_{\rm thermal} = \left(\frac{c_0}{v_{\rm p}}\right)^{3} M_* \, ,
\end{equation}
and $M_{\rm p} \sim 0.56 M_{\rm thermal}$. It is known that in this regime, the flow pattern around the planet begins to deviate from linear calculations \citep[e.g.,][]{Korycansky96}, which may well lead to modifications to the linear torque. To confirm this, we perform the same simulations, but for a 1 $M_\oplus$ planet, which is highly sub-thermal ($M_{\rm p}\ll M_{\rm thermal}$). The results are shown in \figref{fig:1ME}. For this smaller planet, the trend in $\gamma$ is reversed, and becomes compatible with the isothermal linear estimate. \figref{fig:tor_den_1ME} plots the torque density for some of these models, clearly showing the absense of any co-orbital component. The model by \citet{DAngelo10} is also in the highly sub-thermal regime, and we clearly see that their torque density follows the trend in $\gamma$ much better in \figref{fig:tor_den_1ME} than in \figref{fig:tor_den_5ME}. We therefore conclude that the corotation torque in the 5 $M_\oplus$ models is nonlinear in nature.

Corotation torques in inviscid disks are expected to saturate over a few libration times. This cannot be seen with our 5 orbits run time, and much longer simulations are beyond the capacity of our computational resources. We do however note that torque saturation is a prediction for 2D linear flow, and it is unclear whether the same applies for 3D nonlinear flow.

While our two codes are generally in excellent agreement, we do find torque measurements with \texttt{FARGO3D} to have stronger fluctuations then \texttt{PEnGUIn}. This is in line with the more turbulent flow pattern it finds, which we discuss in \secref{sec:flow}. We suspect that they are numerical artifacts generated from the interfaces between different mesh refinement levels. The level of fluctuation is sufficiently small that it does not affect our main conclusions.

Our torque measurements significantly deviate from the one reported by
FAW15, which was $-0.8~\Gamma_0$. In terms of our parameters, their
simulation uses $\{\rs/\rH,~\gamma\}=\{0.1,~1\}$, and a resolution
that translates to about 3 cells per $\rs$. Using our resolution study as a guideline,
it is likely that their resolution was insufficient for a converged torque
measurement. We expect, from extrapolating our results, that the
converged torque measurement in their disk model should be less than
$-2.5~\Gamma_0$ instead. We have attempted to verify this prediction
using our high resolution grid with the same parameters $\{\rs/\rH,~\gamma\}=\{0.1,~1\}$, but the resulting torque strongly fluctuates
between -10 and 5 $\Gamma_0$ \footnote{FAW15's measurement and ours differ in two ways: FAW15 excluded a sphere of 0.5$\rH$ radius around the planet, whereas we made no excision; and FAW15's simulation included a low level of viscosity, while ours is inviscid. Both of these differences stem from FAW15's attempt to reduce numerical noise, which we avoid in order to obtain true numerically converged values.}, which suggests that our resolution is still insufficient. We therefore caution the reader that 3D inviscid isothermal simulations may have a resolution requirement substantially more severe than commonly expected.

For the radiative simulations, more care has to be taken because the 
disk requires more time to adjust thermally. To obtain the final 
torque, we follow a three-steps procedure: first, we perform a 2D ($r-\theta$), 
axisymmetric simulation of the disk (without a planet) to bring the disk 
to thermal equilibirum; second, expand the 2D disk along the azimuth into our 3D low resolution grid, 
introduce the planet and continue the run for 50 orbits; and finally, 
interpolate the grid into medium or high resolution and continue for 
5 additional orbits. \figref{fig:rad} plots the torque in these last 5 orbits.
On the left panel we see that the final torque in high resolution converges 
to a moderately larger value than in low resolution. On the right panel 
we plot the torque from the high resolution
run with two different values of the smoothing length
$\rs$. Same as our isentropic results, there is no clear dependence on the smoothing length. 
Inserting our disk model into the formula of \break
\citet{Masset10}, and using an updated estimate 
of the width of the horseshoe region \citep{Lega15,Masset16} gives a torque value of $0.34~\Gamma_0$, 
in rough agreement with our results. We note that when considering radiative effects, the corotation
torque can be positive and possibly dominate over the negative
Lindblad torque \citep{Masset09,Masset10,Paardekooper10,Paardekooper11}, leading to outwards migration as we observe here.

\subsection{Flow Pattern around the Planet}
\label{sec:flow}

\begin{figure*}[]
\includegraphics[width=1.99\columnwidth]{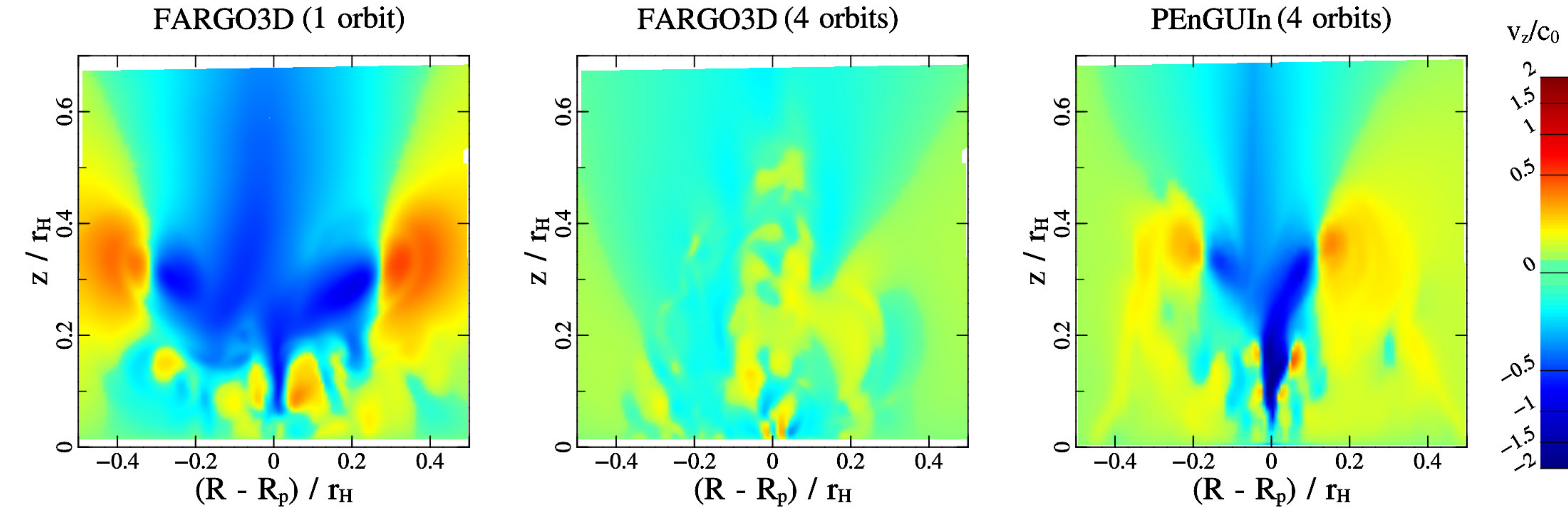}
\caption{Vertical velocity $v_{\rm z}$ at an azimuthal slice across the planet's position. The planet is located at $\{R,~z\}=\{\rp,~0\}$. Red (positive values) indicates velocities upward (away from the midplane), and blue (negative values) indicates downward (toward the midplane). $v_{\rm z}$ is normalized to the initial disk sound speed at the planet's location $c_0$, and the color stretch scales with $\sqrt{v_{\rm z}}$ to emphasize finer details. All panels show the $\{\rs/\rH,~\gamma\}=\{0.03,~1.2\}$ model at high resolution for 5 $M_\oplus$ planets. The left panel is from \texttt{FARGO3D} at 1 orbit, middle panel is the same simulation but at 4 orbits, and right panel is from \texttt{PEnGUIn} at 4 orbits. At 1 orbit, \texttt{FARGO3D} finds a structured flow pattern inside the planet's Hill sphere, with supersonic flow directed toward the planet from above. This pattern breaks down to a flow resembling turbulence at 4 orbits. In contrast, \texttt{PEnGUIn} sustains a structured flow pattern for the entire duration of the simulation.}
\label{fig:flow_far_pen}
\end{figure*}

\begin{figure*}[]
\center
\includegraphics[width=1.4\columnwidth]{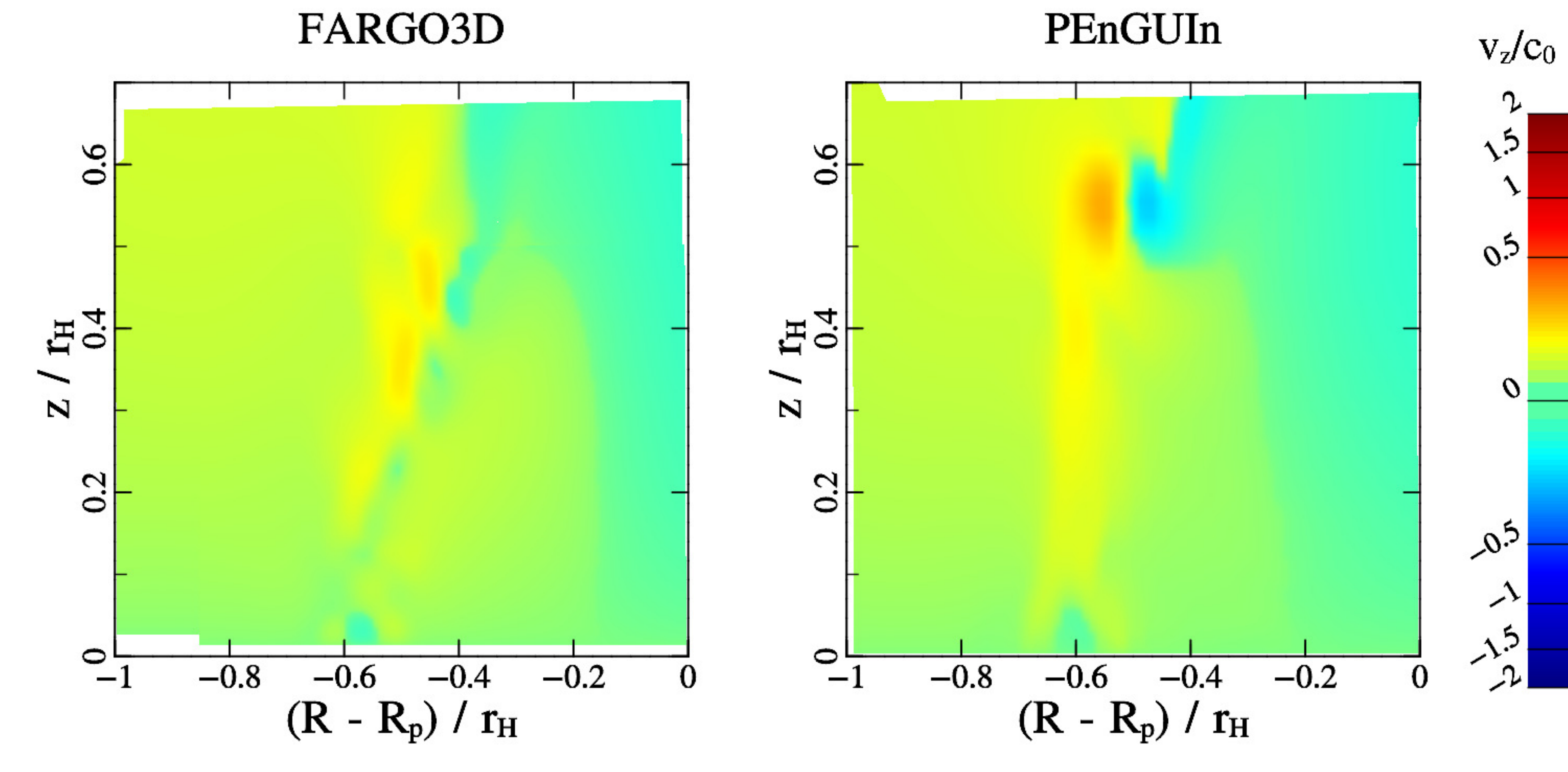}
\caption{Vertical velocity $v_{\rm z}$ plots same as \figref{fig:flow_far_pen}, but at an azimuthal slice offset from the planet's position by 0.01 radian in azimuth. Also note that the x-axis is shifted inward from the planet's orbit. We plot the $\{\rs/\rH,~\gamma\}=\{0.03,~1.2\}$ model at high resolution, obtained at 4 orbits for both \texttt{FARGO3D} (left) and \texttt{PEnGUIn} (right). We find meridional vortices, which are illustrated here by the strong, localized velocity shear, traveling downstream from the planet in both codes.}
\label{fig:flow_downstream}
\end{figure*}

\begin{figure*}[]
\includegraphics[width=1.99\columnwidth]{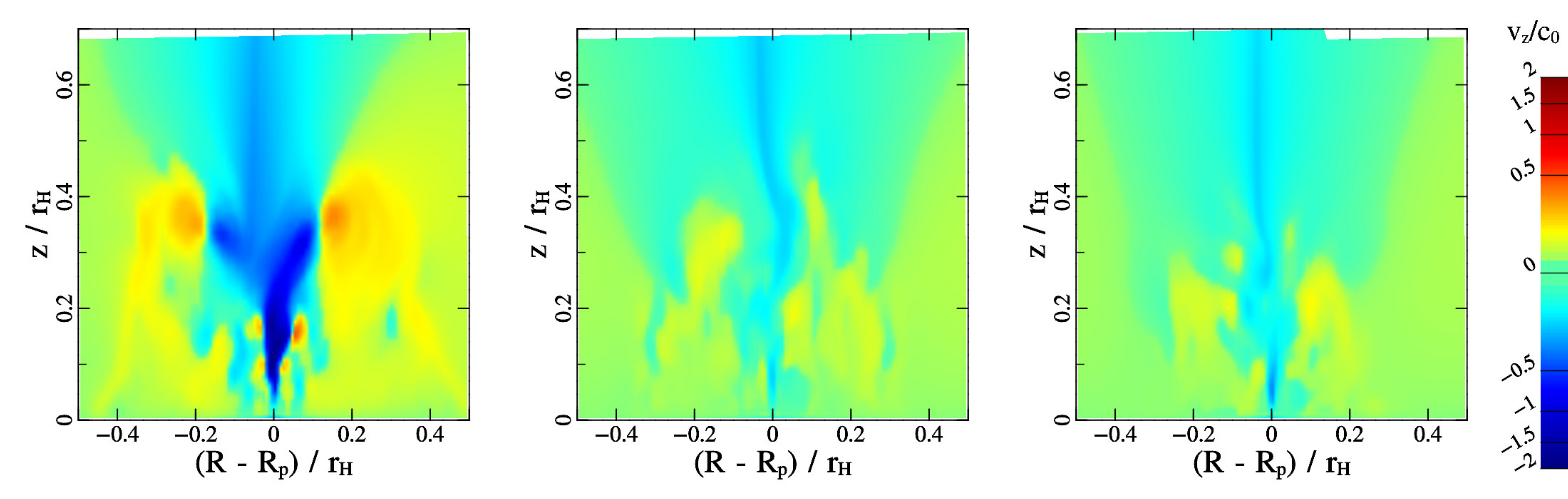}
\caption{Vertical velocity $v_{\rm z}$ plots same as \figref{fig:flow_far_pen}, but with different disk models. The left panel is identical to the right panel of \figref{fig:flow_far_pen}, shown here for comparison. The middle panel plots a model with a larger $\rs$: $\{\rs/\rH,~\gamma\}=\{0.05,~1.2\}$, and the right panel with a larger $\gamma$: $\{\rs/\rH,~\gamma\}=\{0.03,~1.4\}$. All three panels are from \texttt{PEnGUIn} simulations. The maximum vertical flow speed on each panels are 2.0, 0.16 and 0.28 $c_0$ from left to right respectively, indicating that flow speeds are slower when $\rs$ or $\gamma$ is larger.}
\label{fig:flow_rs_gam}
\end{figure*}

In this section we investigate the connection between the flow pattern around the planet and the planetary torque. Previous simulations of embedded planets have seen vertical flow toward the planet's poles (\citealt{Kley01,Klahr06,Tanigawa12,Szulagyi14,Morbidelli14,Ormel15b}; and FAW15), and FAW15 further showed that this results in the generation of meridional vortices. A similar flow pattern is recovered in some, but not all, of our simulations. In particular, we find significant disagreement between \texttt{FARGO3D} and \texttt{PEnGUIn} simulations.

\figref{fig:flow_far_pen} plots the vertical velocity $v_{\rm z}$ in a meridional slice across the planet's location, corresponding to the $\{\rs/\rH,~\gamma\}=\{0.03,~ 1.2\}$, high resolution model. It shows that despite the close agreement in torque measurements between \texttt{FARGO3D} and \texttt{PEnGUIn}, their flow patterns are different. At 4 orbits, \texttt{PEnGUIn} shows an organized flow structure where a fast ($\sim2 c_0$) vertical down flow is directly above the planet, and two vortices are generated at $z\sim0.4\rH$, qualitatively similar to those reported by FAW15 (e.g. their Figure 11). \texttt{FARGO3D} shows a similar pattern only at $t\sim1$~orbit; for $t\ge 2$~orbits, it displays a turbulent flow within the planet's Hill sphere, with a maximum speed of $\sim 0.2 c_0$. This disagreement between the two codes is generally present when $\rs=0.03\rH$ and at high resolution. It is unclear to us which of these flows patterns, turbulent or organized, is correct, but we are encouraged that the torque measurements appear to be robust against these discrepancies.

Despite their differences, both \texttt{FARGO3D} and \texttt{PEnGUIn} find that the planet sheds meridional vortices downstream, in agreement with FAW15. \figref{fig:flow_downstream} again plots $v_{\rm z}$, but in a meridional slice offset from the planet's location by 0.01 radian. The vortex, seen as a strong, localized shear in $v_{\rm z}$, is evident in both codes, with the vortex in \texttt{FARGO3D} being noticeably weaker.

\figref{fig:flow_rs_gam} uses three models from \texttt{PEnGUIn} to show how the flow pattern depends on $\rs$ and $\gamma$. The right panel is identical to the left panel of \figref{fig:flow_far_pen}; the middle panel shows the flow when $\rs$ is larger; and the left panel when $\gamma$ is larger. In general, the flow pattern is preserved when varying $\rs$ and $\gamma$, and only the flow speed changes. Increasing $\rs$ or $\gamma$ both reduces the flow speed; the same applies for \texttt{FARGO3D} simulations as well.

\section{Conclusion and Discussion}
\label{sec:conclude}

We perform hydrodynamical simulations of disk-planet interaction using two different codes, \texttt{FARGO3D} and \texttt{PEnGUIn}, and measure the torque on a planet embedded in a 3D isentropic disk. We find that the torque is independent of the smoothing length $\rs$ (\figref{fig:smooth}), and only weakly dependent on the adiabatic index $\gamma$ (\figref{fig:gam}). In order to obtain convergence on these measurements, we vary the resolution of our simulations, and find that convergence requires at least 3 cells per smoothing length $\rs$ when $\gamma\gtrsim1.4$, and $4\sim5$ cells when $\gamma=1.2$. Overall, our two codes show close agreement. For 5 $M_\oplus$ planets, we observe a nonlinear behavior in the torque, where a weak corotation torque is present despite our constant entropy and vortensity profiles. This corotation torque is more negative when $\gamma$ is smaller, with a magnitude as large as $\sim0.3~\Gamma_0$ when $\gamma=1.2$. For 1 $M_\oplus$ planets, the net torque is more negative with a larger $\gamma$, and no corotation torque is found. The net torque agrees with the results of \citet{Tanaka02} when extrapolated to $\gamma=1$.

To further assess the issue of numerical convergence, we additionally perform simulations with explicit radiative transfer using \texttt{FARGOCA}, and again find good agreement between the measured torques and the analytic formula of \citet{Masset10}. These results confirm what was previously found in lower resolution simulations by \citet{Lega15}.

The flow field around the planet reveals some dependency on $\rs$. In \texttt{PEnGUIn} simulations, the flow pattern around the planet is largely preserved when $\rs$ is varied, but the flow speed decreases with a larger $\rs$. With \texttt{FARGO3D}, we find turbulent flow when $\rs$ is small, and calmer, slower flow when $\rs$ is larger. Increasing $\gamma$ has a similar effect as increasing $\rs$; it also reduces the flow speed.

Because our torque measurements are converged, both between the two codes and in resolution, we are confident in their accuracy. To account for a realistic migration rate, however, a few caveats need to be addressed. First, we measure the planetary torque after the planet has been introduced for 5 orbits, which is a short time compared to the migration timescale of the planet. At low resolution, we extend one of our simulations to 100 orbits, and find that the planet begins to open a gap, consequent of our disk being inviscid. This results in a torque that strongly fluctuates in time. Numerical convergence in this case is difficult to achieve, since long-term high resolution runs are prohibitively expensive. Second, our planet has a fixed, circular orbit. If the planet is allowed to migrate in accordance to the torque it receives, it can potentially modify the torque value. In our current setup, resolution is concentrated at the planet's initial position, so it is not fit to simulate a migrating planet. Future work with a different numerical treatment is required to study this aspect of migration. Finally, our parameter space is restricted to disks with an initial zero vortensity gradient. Allowing for a finite gradient is likely to produce a larger horseshoe drag that can alter the magnitude, or even the direction of migration. While this type of study has been done previously in 2D \citep[e.g.,][]{Casoli09, Masset09}, detail 3D studies have only been done for highly sub-thermal planets ($M_{\rm p}\ll M_{\rm thermal}$) \citep{Masset16}. The meridional vortices observed here and in previous work may play a role in the 3D horseshoe drag for near to super-thermal planets ($M_{\rm p}\gtrsim M_{\rm thermal}$). Indeed, we have uncovered a 3D corotation torque for nearly-thermal planets in this study. Its magnitude appears small compared to the net torque, but it is possible that our particular choice of a zero vortensity gradient minimized it.

While our torque measurements appear robust, the flow structure around the planet shows discrepancies between our two codes. Moreover, the flow is sensitive to $\rs$, which controls the flow speeds within the Hill sphere, or even generate turbulence in $\texttt{FARGO3D}$ simulations. The flow speed is important for determining how fast the gas within the Hill sphere circulates with the rest of the disk, and has implications on the thermal cooling of the planet's atmosphere \citep{Ormel15b}. Additionally, our result implies that $\rs$ can potentially affect numerically measured accretion rates, such as those by \citet{Machida10} and \citet{Szulagyi14}. It remains to be seen whether a sufficiently small $\rs$ will produce a converged flow structure. Further code testing is also needed to understand the difference between our two codes.

\acknowledgments We thank an anonymous referee for helpful feedback that improved this manuscript. JF gratefully acknowledges support from the Natural Sciences and Engineering Research Council of Canada, the Center for Integrative Planetary Science at the University of California, Berkeley, and the Sagan Fellowship Program under contract with the Jet Propulsion Laboratory (JPL) funded by NASA and executed by the NASA Exoplanet Science Institution.
FM and DV gratefully acknowledge support from UNAM's DGAPA grant PAPIIT~IN101616, and support from CONACyT grant~178377.
EL is thankful to ANR for supporting the MOJO project
(ANR-13-BS05-0003-01).
This work was performed using HPC resources from GENCI [IDRIS]
(Grant 2016, [i2016047233]). 

\bibliographystyle{apj}
\bibliography{Lit}

\end{document}